\documentclass[12pt]{iopart}
\usepackage{txfonts}
\usepackage{graphicx}
\usepackage{color}

\begin{document}

\title{(In)finite extent of stationary perfect fluids in Newtonian theory}

\author{Patryk Mach$^1$ and Walter Simon$^{1,2}$}

\address{$^1$ M.~Smoluchowski Institute of Physics, Jagiellonian University, Reymonta 4, 30-059 Krak\'{o}w, Poland}
\address{$^2$ Gravitational Physics, Faculty of Physics, Vienna University, Boltzmangasse 5, A-1090 Vienna, Austria}
\eads{\mailto{mach[at]th.if.uj.edu.pl},~~\mailto{walter.simon[at]univie.ac.at}}
\begin{abstract}
For stationary, barotropic fluids in Newtonian gravity we give simple criteria on the equation of state and the 
``law of motion'' which guarantee finite or infinite extent of the fluid region
(providing {\it a priori} estimates for the corresponding stationary Newton--Euler system). Under more restrictive conditions, we can also exclude the presence of 
``hollow'' configurations. Our main result, which does not assume axial symmetry, uses the  virial theorem
as the key ingredient and generalises a
known result in the static case. In the axially symmetric case  stronger
results are obtained and examples are discussed.
   
\end{abstract}


\section{Introduction}

This work deals with a priori estimates of solutions of the stationary Newton--Euler system of equations. For definiteness the latter includes 
a barotropic equation of state (EOS) $\rho_g = \rho_g(p)$ relating the matter density and the pressure, and a ``law of motion'' (LOM) specifying 
the velocity $\vec{U}(x^i)$ or the centrifugal potential  $\phi_c = \phi_c(x^k)$ as a function of position. This system can be used to model stars or 
galaxies, provided the perfect fluid is a viable model for the corresponding multi-particle system consisting of molecules or stars, respectively. 
The equations have been studied accordingly, from mathematical as well as from physical viewpoints, see
e.g.~\cite{AK}--\cite{BT}. 
Key problems are existence, uniqueness, (axial) symmetry, stability and parametrisation of
the solutions, where uniqueness may be understood modulo global parameters, like mass and angular momentum. Other useful 
parametrisations are the pressure or the density at the centre or at the the axis, depending on the symmetry, 
or, as we shall see below, the sum of the gravitational potential
$\phi_g$ and the centrifugal potential $\phi_c$ on the surface,  called $\Phi_S$.      

In the present work we assume a connected fluid region and focus on investigating the problem if the Newton--Euler system admits (only) solutions of 
finite or infinite extent (i.e., compact support of the density function). In fact this question can be regarded as a 
special case of the problem of obtaining a formula for $\Phi_S$ or a bound thereon, for which we give results as well. 
Normally finite extent of solutions is a prerequisite for the physical significance of a model.
This always applies to models for single stars. However, for the polytropic EOS
\begin{equation}
\label{pol}
 p  =  \frac{1}{n + 1}\rho_o^{-\frac{1}{n}}~ \rho_g^{\frac{n+1}{n}}, \qquad {\rho_o = \mathrm{const.} > 0}
\end{equation}
 with index $n = 5$, all static configurations are infinitely extended. They are known as ``Plummer's model'' for non-rotating globular clusters of stars (see e.g., \cite{BT}).

In the static case, systematic treatments of the finiteness problem, which do not rest on the assumption of spherical symmetry,
can be found in \cite{WS1,LM,WS2}; for stronger results in spherical
symmetry, cf.~\cite{MH}. Formally, the task consists of determining if the surface value $\Phi_S$ of the gravitational potential $\phi_g$ agrees with 
its value at infinity, which we set to zero. Physically, the key ingredient is the continuum form of the virial theorem,
and mathematically it is a modified Pohozaev--Rellich identity. Moreover, since $\rho_g\ge 0$, the maximum principle is available as well.
In \cite{WS1,WS2} the functional
\begin{equation}
F(p) =  \rho \int_0^p \frac{dp'}{\rho_g(p')} - 6p \label{F}
\end{equation}
was identified as the crucial quantity in the sense that $F \le 0$ for all $p$ guarantees finiteness of the solutions unless $F \equiv 0$, which  
characterises the polytropes of index 5. Moreover, $F \ge 0$ for all $p$ but $F \not\equiv 0$ implies that there are no solutions 
with finite mass. If $F$ changes sign, the analysis of the spherically symmetric Newton--Euler system is much more involved and uses dynamical systems 
techniques (see e.g., \cite{HU}).

In the stationary case, results on finiteness of which we are aware are basically amendments to theorems on existence (cf.~the classic paper 
\cite{AB}, the recent account \cite{LS} and the references therein). All these results require axial symmetry and  conditions on the EOS which are in many 
respects more restrictive than ours. In contrast, the present paper ignores the problem of existence, which means that we are interested in ``a priori'' 
estimates. On the other hand, some of our results do not require axial symmetry, which does not hold in general for stationary perfect fluids 
(cf. Sect.~3 below) and we use appropriately 
adapted conditions on the EOS and the LOM. Formally, we compare now the value of the ``effective potential'' $\phi = \phi_g + \phi_c$ at infinity 
(where we can again set it to zero unless it diverges) and at the surface ($\Phi_S$), and the main technical tool is still the virial theorem.  The 
quantity characterising rotation which arises in our analysis is   
\begin{equation}
D = x^i \frac{\partial}{\partial x^i} \phi_c + \frac{1}{2} \phi_c,
\label{D}
\end{equation}
where $x^i$ denote Cartesian coordinates. Finiteness of the solution is guaranteed provided $D$ and $F$ have the same 
sign everywhere but do not both vanish identically. Under the additional
requirement that $\Delta \phi_c \ge 0$ the case with $D$ and $F$ 
being non-negative can be excluded by the maximum principle, again unless both $F$ and $D$ vanish everywhere. 
This latter case leads to a particular LOM for the $n = 5$ polytrope, which
will be analysed separately. 

For polytopes with index $n$ which rotate according to a ``power law''  $\vec{U}
\propto r^{-m} \partial/\partial \varphi$, where $r$ is the distance from an 
axis and $m \in \mathbf{R}$, our conditions $F\le 0$ and $D \le 0$ enforcing finiteness read $n \le 5$ and $m \ge 5/4$, respectively. In particular, a 
polytropic fluid  with $n \le 5$  whose layers rotate at or near their Kepler orbits $(m \simeq 3/2)$ must be finite. This is 
somewhat counter-intuitive for the following reason: For such a fluid the gravitational attraction
between its ``layers'' will be balanced locally only by the centrifugal
force and not by pressure, whence the fluid should behave like ``dust''. However, there is no obvious reason why Keplerian orbits
of dust (-particles) should not extend to infinity. Of course this
``paradox'' must disappear upon properly taking into account the gravitational interaction between the 
dust particles or fluid layers. However, this reminds of heuristic arguments trying to estimate the amount of dark matter necessary to 
stabilise galaxies 
and to explain the observed rotation law $m \approx 1$. Recent analyses of the dark matter problem by different methods (see e.g., \cite{JBK} and 
\cite{DS}) indicate that for this purpose much less dark matter is needed than previous approximations suggested. In resolving this issue Vlasov--Poisson 
theory should play a key role and also establish a connection to the phenomena in fluid mechanics described here (see e.g., \cite{RR}).

While we believe that our results are new, they are rather elementary, and the axially symmetric ones might well be contained in 
the vast astronomical, physical and mathematical literature on rotating fluids since Euler's and Newton's time. On the other hand, our exposition 
is motivated by and partially adapted to the corresponding relativistic problem. 
In fact for static perfect fluids in general relativity the quantity $F(p)$
in (\ref{F}) can be replaced by 
 \begin{equation}
G(p) =  \rho \left[ \exp \left( \int_0^p \frac{dp'}{\rho_g(p') + p} \right) - 1
\right] - 6p \label{G}
\end{equation}
to obtain analogous conclusions regarding finiteness and infiniteness as in the Newtonian case \cite{WS1,WS2}. 
However, since the known relativistic  virial theorems \cite{GC,SBEG} are not suitable for
the present purpose the methods are different, and generalisations to the stationary case are not 
straightforward. We intend to present relativistic analogues of some of the results given below elsewhere. 

\section{Assumptions and Basic results}

This section is divided into five subsections. In Sect.~2.1 we give some basic definitions and continue in  Sect.~2.2 with writing the 
Newton--Euler system in a form suitable for our purposes. Sect.~2.3 contains a discussion of the  EOS $\rho_g = \rho_g(p)$
 and the relationships  $\rho_g(\phi)$ and $p(\phi)$. 
(The symbols were defined in the introduction).  The space-time dependence of the gravitational variables
$\rho_g$, $\phi_g$  and the rotational variables $\rho_c$, $\phi_c$ will be discussed in Sects.~2.4 and 2.5, respectively, 
and the latter section also contains the key Lemma 2.5.2 on (in)finiteness. 

We denote Cartesian coordinates by either $x^i$ $(i = 1,2,3)$ or by $x,y$ and $z$.  
$R = \sqrt{x^2 + x^2 + z^2}$, $\vartheta$ and $\varphi$ are spherical polar
coordinates, and  $r = \sqrt{x^2 + y^2}$ and $\varphi$ are cylindrical polar
coordinates. The volume element on $\mathbf{R}^3$ is denoted by $d\nu$.

\subsection{Finiteness versus infiniteness}

The vacuum and the fluid regions  ${\cal V}$ and ${\cal F}$ are {\it by
definition} 3-dimensional, open sets where $\rho_g = 0$ and $\rho_g \neq 0$,
respectively. This means that all points $v \in {\cal V}$ and $f \in {\cal F}$
have open 3-neighbourhoods in ${\cal V}$ and ${\cal F}$.
 We take the ${\cal F}$ to be connected but possibly with non-trivial
topology; in particular ${\cal V}$ and ${\cal B} = \partial {\cal F}$ may be disconnected. 
The boundary ${\cal B}$ is required to be a $C^1$ submanifold with ${\cal F}$ lying only on one side of 
${\partial \cal F}$. 
In other words,  $\rho_g$ only vanishes in vacuum and possibly at the boundary, but not
on sets of dimension less than three ``inside the region occupied by the
fluid.'' 
The reason for this requirement is to avoid trouble with integrating Euler's
equation (\ref{eul}).
We also note that the boundary can be characterised by vanishing pressure, cf.~Sect.~2.3.  

The following definition introduces a shorthand for our main issue. 

\vspace{1ex} \noindent \textit{Definition 2.1.} The fluid region ${\cal F}$ is called finite if it stays within a compact subset of
$\mathbf{R}^3$, and infinite otherwise. 

\vspace{1ex} In particular, fluids which extend to infinity in at least one direction are called infinite.
In the axially symmetric case dealt with in Sect. 3 we will distinguish between
(in)finite extent in axial and equatorial directions.

\subsection{The Newton--Euler system}

We denote by  $\vec{U}$ the velocity of the fluid and by 
${\cal I}$ the integral
\begin{equation}
\label{I} 
{\cal I} = \int_0^p \frac{dp'}{\rho_g(p')}
\end{equation}
(the specific enthalpy), whose existence is assumed for finite $p$; this is
satisfied in particular for polytropes $\rho_g(p) \propto p^a$ when $a < 1$. 
Further restrictions on the EOS are discussed in Sect.~2.3.
The stationary Newton--Euler system can be written as follows
\begin{eqnarray}
\rho_g & = & \rho_g(p), \label{eos}\\
\Delta \phi_g & = & 4 \pi \rho_g,  \label{new}\\
\nabla \left(\rho_g \, \vec{U} \right)  & = & 0, \label{con} \\
 - \left(\vec{U}.\nabla \right) \vec{U} & = & \nabla \phi_g + \frac{\nabla p}{\rho_g}  =  \nabla \left(\phi_g + {\cal I} \right). \label{eul}
\end{eqnarray}

In terms of the centrifugal potential $\phi_c$ defined up 
to a constant (which will be specified in Sect. 2.5) by
\begin{equation}
 \left(\vec{U}.\nabla \right) \vec{U}   =  \nabla \phi_c, \label{centr_pot}
\end{equation}
(\ref{eul}) yields the ``Bernoulli'' equation
\begin{equation}
\phi_g + \phi_c + {\cal I}(p)  =  \Phi_S,   \label{ber} 
\end{equation}
where $\Phi_S$ is a constant. We also introduce a ``centrifugal charge
density''
\begin{equation}
\label{in}
\rho_c =   \frac{1}{4 \pi} \mbox{div} \left[ \left(\vec{U}.\nabla \right) \vec{U} \right],
\end{equation} 
in terms of which Eq.~(\ref{eul}) yields
\begin{equation}
 \Delta \phi_c  =  4 \pi \rho_c.  
\label{poi}  
\end{equation} 

\subsubsection*{Remarks.}
\begin{enumerate}

\item The existence theorems in the axially symmetric case show that the motion of the fluid has to be specified somehow in the 
Newton--Euler system. 
A simple way of doing so is to prescribe  the velocity in terms of position $\vec{U} = \vec{U}(x^i)$ or to prescribe the centrufugal 
potential $\phi_c(x^i)$ which we do below (without restriction to axial
symmetry).  
An alternative is to specify the angular momentum per unit mass (see e.g., \cite{AB}). 

\item The above definitions of $\vec{U}$, ${\phi_c}$ and ${\rho_c}$, and the
equations containing them, are understood to hold only in ${\cal F}$. 
It will however be convenient to extend these quantities, and some relations
between them, to ${\cal V}$. In Sect.~2.5 we will discuss four alternative extensions which
will be used in the following sections.

\item
In work focusing on existence, the aim is to specify only the EOS and the velocity or the specific angular momentum as functions in space, 
and to get information on the spatial behaviour of all variables. In this respect the present work has the same scope, although existence 
is not the issue here. However, except for crucial conditions on the EOS and the LOM, we will
in the sequel also have to make differentiability
and falloff requirements for the space-time dependence of our functions.
\end{enumerate}
     
\subsection{The Equation of state and the effective potential}

The ``effective potential'' is defined by $\phi = \phi_g + \phi_c$;  its gradient is called ``effective gravity'' in Sect.
3.2.1 of \cite{JT}.

Our first Lemma serves mainly to list the assumptions on the equation of state required
later. The proof is an easy consequence of the Bernoulli equation (\ref{ber}).

\vspace{1ex} \noindent \textit{Lemma 2.3.1.} We assume that  $\rho_g(p)$ is piecewise continuous,  $0 \le \rho_g(p) < \infty$, and that 
the integral ${\cal I}$ exists for finite $p$. 
Then in ${\cal F}$ the effective potential $\phi(p)$ is $C^0$, piecewise $C^1$, and strictly monotonic; the same applies to the 
inverse $p = p(\phi - \Phi_S) = p(\phi)$, and the density $\rho_g$ is also a $C^0$ and piecewise $C^1$ function of $\phi$ that satisfies 
$\rho_g = d p(\phi)/d\phi$.

\subsubsection*{Remark.}  The Lemma implies that the surface $p = 0$ is an equipotential surface of $\phi$, and $\phi(p = 0)$ takes the 
value $\Phi_S$. If we allowed for disconnected fluid regions, a consistent definition of $\phi$
on $\mathbf{R}^3$ (cf. Sect. 2.5) would imply different constants $\Phi_S$ on each component in general. 

\vspace{1ex} The next Lemma (which is known, see e.g., \cite{BS}) contains a stronger assumption on the EOS in a neighbourhood of $p=0$, 
which will also be made in Proposition 4.2.

\vspace{1ex} \noindent \textit{Lemma 2.3.2.} In addition to the requirements
of Lemma 2.3.1, we assume that $\rho_g(p)$ is $C^0$ in $[0,\delta)$ and
$C^1$ in $(0,\delta)$ for some $\delta > 0$.
Then $\lim_{p \rightarrow 0} p/\rho_g(p) = 0$.

\vspace{1ex}\noindent \textit{Proof.} The result is obvious if $\rho_g(0)
\neq 0$. If $\rho_g(0) = 0$ we first note that near $p=0$ the inverse $p= p(\rho_g)$ exists, and $dp/d\rho_g \ge 0$. 
We can thus replace the assertion by  $\lim_{\rho \rightarrow 0}
p(\rho)/\rho_g = 0$ (dropping the subscript $g$ on $\rho_g$ when the latter is a sub- or superscript
itself).
We obtain
\begin{eqnarray}
\fl \infty  & > &  \lim_{\epsilon \rightarrow 0} \int_{\epsilon} ^p \frac{dp'}{\rho_g(p')} = 
\lim_{\epsilon \rightarrow 0} \int_{\rho(\epsilon)}^{\rho_g(p)}
\frac{dp'}{d\rho_g'} \frac{d\rho_g'}{\rho_g'} 
\ge \lim_{\epsilon \rightarrow 0} 
\left\{ \left[ \inf_{[\rho_g(\epsilon), \, \rho_g(p)]} 
 \frac{dp(\rho_g)}{d\rho_g} \right] \int_{\rho(\epsilon)}^{\rho(p)}
 \frac{d\rho_g'}{\rho_g'}  \right\} \ge  \nonumber\\
\fl {} & \ge &  \lim_{\rho \rightarrow 0} \frac{dp(\rho_g)}{d\rho_g}  \lim_{\epsilon \rightarrow 0} 
 \int_{\rho(\epsilon)}^{\rho(p)} \frac{d\rho_g'}{\rho_g'} = 
\lim_{\rho \rightarrow 0} \frac{dp(\rho_g)}{d\rho_g}~ .~ \infty.
  \end{eqnarray}
Hence $0 = \lim_{\rho \rightarrow 0} dp(\rho_g)/d\rho_g = \lim_{\rho \rightarrow 0} p(\rho_g)/\rho_g$
 by de l'Hospital's rule.

\subsection{The gravitational variables}

We work in the weighted Sobolev spaces  $W^{k,p}_{\delta}$ 
($1\le p \in {\mathbf R}$,
$\delta \in {\mathbf R}$, $k \in {\mathbf N}_0$), based on the weighted
Lebesgue norms
\begin{equation}
||u||_{k,p,\delta}  =  \sum_0^k ||D^ju||_{p,\delta - j}, \qquad
||u||_{p,\delta} = \left( \int_{{\mathbf R}^3} |u|^{p} \sigma^{-\delta p - 3} d \nu
\right) ^{1/p}
\end{equation}
for measurable functions $u \in L^p_{loc} \left({\mathbf R}^3 \right)$,
where $\sigma = (1 + R^2)^{1/2}$.
This is Bartnik's index convention, cf.~\cite{RB}, see also remark (ii) below.

We always require that the fluid has finite mass $m = \int_{\mathbf R^3} \rho_g d\nu$. 
The following Lemma is standard; we use \cite{MO,RB} for the inversion of the Laplacian
in (\ref{new}), and the strong maximum principle, Thm.~9.6 of \cite{GT}. 

\vspace{1ex} \noindent \textit{Lemma 2.4.} Let $ \rho_g  \in W^{0,2}_{-3-\alpha}$,
$0 < \alpha < 1$.
Then there is a unique solution $\phi_g \in W^{2,2}_{loc}$ of (\ref{new}) with 
$\psi_g =  \phi_g - m/\sigma \in W^{2,2}_{-1-\alpha}$. Moreover, $\phi_g \le 0$.

\subsubsection*{Remarks.}

\begin{enumerate}
\item
Weighted Sobolev spaces have the important property that a rather slow falloff of the density function in one or two 
directions is admitted as long as it is compensated by sufficiently fast falloff in the other direction(s). In particular, an axially 
symmetric disk of finite thickness with a measurable density function  $\rho_g \in W^{0,2} = L^2 $  and falloff 
$\rho_g = O(1/r^{2+\epsilon})$, $\epsilon > 0$, satisfies the requirement of Lemma 2.4 and of the subsequent results. 
\item
With the conventions of Bartnik \cite{RB} used above, the index $\beta$ for $f \in W^{k,p}_{\beta}$ is related to the 
growth of $f$ at infinity; in particular $f = o(R^{\beta})$ provided $kp > 3$ (throughout the paper, falloff conditions are always understood for large $r$ or $R$). In fact for the function $\psi_g$ 
introduced in Lemma 2.4 it follows that  $\psi_g = o(R^{-1-\epsilon})$. However, getting the corresponding falloff 
for the derivatives, namely $\nabla_i \psi_g = o(R^{-2-\epsilon})$ would require $p>3$. While such
first derivatives do occur  in  Theorem 4.2 below, less precise information on their falloff, which follows from $p=2$, will suffice. 
\end{enumerate}

\subsection{The rotational variables}

As already mentioned in the introduction, the strategy of our finiteness argument is to compare the value of the
effective potential $\phi = \phi_g + \phi_c$ at the surface with its value at infinity. On the other hand, integrating 
Euler's equation (\ref{eul}) defines $\phi_c$ only in ${\cal F}$, and in fact only up to a constant.    

We therefore extend now $\phi_c$ to ${\cal V}$. Below we consider four alternative definitions of the 
rotational variables valid on $\mathbf{R}^3$, labelled A, B$_<$, B$_>$ and C.
 In extensions B$_<$, $B_>$ and $C$ we follow standard practice and
 prescribe $\phi_c(x^i)$ a priori on $\mathbf{R}^3$, 
irrespective of the fluid distribution. Of course we can also  prescribe $\vec U(x^i)$ instead of
 $\phi_c(x^i)$, together with a constant 
in the resulting  $\phi_c$. On the other hand, extension A is motivated by the analogy between the Poisson equations
 (\ref{new}) and (\ref{poi}), which suggests treating  $\rho_c$ as a ``source'' for $\phi_c$. 

While A, B$_<$ and B$_>$ are compatible with any symmetry, definition C refers only to a cylindrically symmetric velocity 
distribution $\phi_c(r)$.
All definitions are in fact not only extensions  from ${\cal F}$ to ${\cal V}$, but involve also extra conditions on the 
falloff of the velocity at infinity if the fluid spreads there (cf.~the remarks after Definition 2.5 for details).

\vspace{1ex} \noindent \textit{Definition 2.5.} For a solution of (\ref{eos}--\ref{ber})
with ${\cal I}$  finite, 
$\rho_g$ and $\phi_g$  as in Lemma 2.4, and $1 < q < \infty$, $0 < \epsilon < 1$,   
we assume one of A, B or C:

\begin{description}
\item[A:] $\rho_c \in W^{0,q}_{-2-\epsilon}$ in ${\cal F}$, $\rho_c \equiv 0$ in ${\cal V}$ and 
$\phi_c \rightarrow 0$ at infinity.
\item[B:] \,\,$\phi_c(x^i) \in W^{2,q}$ given on $\mathbf{R}^3$ such that,
for all radial unit vectors $\vec{n}$ there is a unique limit
\begin{equation}
\label{philim}
\lim_{R~ \rightarrow \infty} \phi_c(R ~\vec{n}) = \phi_{\infty}(\vec{n}) =  \phi_{\infty}(\vartheta,\varphi)
\end{equation} 
and either
\begin{description}
\item[B$_<$:]  $\sup_{\vartheta,\varphi} \phi_{\infty}(\vartheta,\varphi) = 0$, or 
\item[B$_>$:]  $\inf_{\vartheta,\varphi} \phi_{\infty}(\vartheta,\varphi) = 0$. 
\end{description}
\item[C:] $\phi_c(r) \in  C^1(\mathbf{R} \setminus \{0\})$ given (and possibly divergent at $0$), 
and  $\lim_{r \rightarrow \infty} \phi_c(r) = 0$.
\end{description}

\subsubsection*{Remarks.}
\begin{enumerate}
\item While condition A is somewhat alien to the Newtonian case, it is mainly motivated by Relativity. 
There the norm and twist potentials of the stationary Killing field satisfy elliptic equations, 
while on the other hand counterparts of the coordinate conditions B and C will hardly make good sense.  

\item We note that cylindrically symmetric potentials
$\phi_c(r)$ do ${\it not}$ satisfy (\ref{philim}) unless $\phi_c(r) = \mathrm{const.}$,
since the limit $\lim_{R \rightarrow \infty} \phi_c(r) = \lim_{z \rightarrow \infty} \phi_c(r) = \phi_c(r)$ in the axis direction depends manifestly on $r$, while there 
should be a unique $\phi_{\infty}(\vartheta = 0, \phi)$. On the other hand, condition B is meaningful in particular 
for ``almost spherical'' velocity distributions.

\item Since $\phi_c$ was defined in (\ref{ber}) only up to a constant, condition
B amounts to requiring (\ref{philim}), and that
 either  $\sup_{\vartheta,\varphi} \phi_{\infty}(\vartheta,\varphi) < \infty$, or   
$\inf_{\vartheta,\varphi} \phi_{\infty}(\vartheta,\varphi) > - \infty$,
 while C includes the  requirement that 
$\lim_{r \rightarrow \infty} \phi_c(r) > -  \infty$. This latter requirement can in fact be removed, so that $\phi_c(r)$ is
 allowed to diverge both at the axis and at infinity. However, in order not to overload the subsequent definitions and 
results, this option will be considered only in the final remark of Sect.
 3.2 and in examples.
\item
The falloff conditions B$_<$ and B$_>$ and C are in some sense less restrictive than A.
However in Theorem 4.2 the former requirements need to be explicitly supplemented by a falloff condition on the pressure
which, on the other hand, follows from A automatically by virtue of the Bernoulli equation.

\item
None of the definitions A, B$_<$, B$_>$ or C extends the validity of the Bernoulli equation from ${\cal F}$ to the vacuum 
region ${\cal V}$ in general. Such an extension could simply be afforded by setting $\phi_c = -\phi_g + \Phi_S$ in ${\cal V}$, 
but it seems to be of little use. Note in particular that such a $\phi_c$ would not necessarily be $C^1$ at the surface.    

\item In principle, we could also consider axially symmetric velocity distributions of the form $\phi_c(r,z)$. However, 
the Poincar\'{e}--Wavre theorem, Lemma 3.1, then already implies cylindrical symmetry in ${\cal F}$, so an extension 
to ${\cal V}$ with the same symmetry is the natural choice. 

\end{enumerate}

The following Lemma is analogous to Lemma 2.4 but, compared to the gravitational variables, the falloffs are slower
here, and in contrast to the natural condition $\rho_g \ge 0$, the assumption $\rho_c \ge 0$ is highly restrictive.

\vspace{1ex} \noindent \textit{Lemma 2.5.1.} If  $ \rho_c  \in W^{0,2}_{-2-\alpha}$, $\alpha > 0$,
then $\phi_c$ can be chosen such that $\phi_c \in W^{2,2}_{-\alpha}$.
Moreover, if $\rho_c \ge 0$, then $\phi_c \le 0$, and therefore also $\Phi_S \le
0$.

We now have the following easy to prove, but important Lemma.

\vspace{1ex} \noindent \textit{Lemma 2.5.2.}  Under the requirements and with
the labelling of Definition 2.5 the following holds:
\begin{description}
\item[A:] If $\Phi_S \ne 0$ then the fluid is finite.
 Moreover, if $\rho_c \ge 0$ the fluid is finite if $\Phi_S < 0$ and
 infinite if $\Phi_S = 0$. 
\item[B:] If $\phi_{\infty}(\vartheta,\varphi)$ does not agree with $\Phi_S$ for some $(\vartheta, \varphi)$, the fluid is finite 
``in the direction $(\vartheta,\varphi)$,'' or more precisely, it does not intersect any 2-sphere
$\mathbf{S}^2_R$ of sufficiently large radius
$R$ at the points $(R,\vartheta,\varphi) \in \mathbf{S}^2_R$.  In particular, for extensions B$_<$, B$_>$ the fluid is finite if $\Phi_S > 0$, 
$\Phi_S < 0$ respectively.
\item[C:] If $\Phi_S < 0$, the fluid is finite (in all directions), and if $\Phi_S > 0$ the fluid is finite in every direction except possibly 
in the axial one.  Moreover, if $\Phi_S = 0$, and the fluid extends to infinity in the axis
direction at some radius $r_1$, it is static for all $r \ge r_1$.  
\end{description}

\vspace{1ex} \noindent \textit{Proof.} 
The proofs of parts A and B follow easily from the definitions, and the second part of case A from the maximum principle. 
The proof of  C is postponed to  Section 3.2.

\vspace{1ex} \noindent \textit{Remark.} We recall from a previous remark that a cylindrically symmetric $\phi_c(r)$ considered under C 
in Lemma 2.5.2 is not compatible with B, in particular not with (\ref{philim}). This accounts for differences in the 
conclusions and the proofs in the corresponding parts of Lemma 2.5.2.

\section{Results with Symmetry}

Assumptions and results on symmetry of barotropic perfect fluids require a careful discussion (cf.~\cite{LL} where the relativistic case is included as well). 

It is important to distinguish between the symmetry of the velocity field $\vec{U}(x^i)$ and its trajectories on the one hand, 
and the symmetry of the whole configuration on the other hand. In fact, a well known example important in theory are non-axially 
symmetric fluid trajectories winding around an axially symmetric torus (cf.~\cite{AK} in the Newtonian case and \cite{BG} in Relativity). 
Fluids with non-axially symmetric configuration have been discussed; cf.~\cite{LL} and Sect.~2.8.3 of \cite{JT}.
On the other hand, known existence proofs of reasonable generality all assume axially symmetric velocities (cf.~\cite{LS} and the references therein).

In the Newton--Euler system (\ref{eos}--\ref{eul}) the gravitational and velocity variables can enjoy different symmetries, as long
as the coupling between the potentials via the Bernoulli equation
(\ref{ber}) is respected. In this work we do not assume any symmetry of the matter variables $\rho_g$ and
$\phi_g$. As to the rotational 
variables we first state in Sect.~3.1 the definitions of axial and cylindrical
symmetry. Then we continue with a known result  (the ``Poincar\'{e}--Wavre theorem'') relating 
them, and give a simple application. In Sect.~3.2 we examine systematically the results arising from the integrals of the cylindrically symmetric 
Bernoulli equation via the limits $r \rightarrow \infty$ and  $z \rightarrow
\infty$. In particular, Sect.~3.2 also contains the proof of Lemma 2.5.2.C.

\subsection{The Poincar\'e--Wavre theorem}

\subsubsection*{Definition 3.1.}
This refers to solutions to the Newton--Euler system (\ref{eos}--\ref{eul}):
\begin{enumerate}
\item A solution has axially symmetric velocity if the velocity $\vec{U}$ is proportional to the axial Killing vector,
and if $|\vec{U}|$ (and hence also $\phi_c$) are rotation invariant.
\item An axially symmetric  solution has cylindrically symmetric velocity 
 if $\vec{U}$, (and hence also $\phi_c$) are invariant under translations
 along an axis.
\end{enumerate}

\vspace{1ex} \noindent \textit{Lemma 3.1.}
For a solution of Newton--Euler system (\ref{eos}--\ref{eul}) with axially symmetric
centrifugal potential $\phi_c(r,z)$,  the latter is in fact cylindrically symmetric, viz. 
\begin{equation}
 \phi_c(r,z) = \phi_c(r), \qquad \vec{U} = \omega(r) \frac{\partial}{\partial \varphi},
\end{equation}
and it holds
\begin{equation}
\label{phic}
\phi_c(r) = - \int^r_{r_1} \omega^2(r^\prime) r^\prime dr^\prime, \qquad r >
r_1 = \mathrm{const.}
\end{equation}

\vspace{1ex} \noindent \textit{Proof.} 
The proof is obvious from Euler's equation which reads
\begin{equation}
\frac{\partial \phi_c(r,z)}{\partial z} = 0, \qquad \frac{\partial \phi_c(r,z)}{\partial r} = - \omega^2(r,z) r  
\end{equation}
under the stated assumptions. 

Here is a simple application of the previous Lemma.

\vspace{1ex} \noindent \textit{Proposition 3.1.} 
We assume a cylindrically symmetric velocity distribution and an EOS as in Lemma
2.3.1. If the fluid extends to infinity in the axis direction at two different radii $r_1 \ge 0$ and
$r_2 > r_1$, the intermediate region 
 ${\cal A} = \{(r,z) ~\mbox{with} ~r_1 < r < r_2  ~\mbox{and}~ -\infty \le z \le \infty \}$ must be static, 
i.e., $\phi_c = \mathrm{const.}$ in ${\cal A}$.

\vspace{1ex} \noindent \textit{Proof.} 
From the Bernoulli equation we obtain 
\begin{equation}
\phi_c(r_1) =  \lim_{z \rightarrow \infty} \phi_c (r_1) =  \lim_{z \rightarrow \infty} 
\left( \Phi_S - \phi_g - {\cal I} \right)=  \Phi_S,
\end{equation}
and analogously  $\phi_c(r_2) = \Phi_S$, which contradicts $(\ref{phic})$,  
unless $\omega \equiv 0$, and the solution is static. 

\vspace{1ex} \noindent \textit{Remark.}
This result implies in particular that solutions with axially symmetric velocity
distribution which fill the whole space must be static and hence spherically
symmetric.  

\subsection{Finiteness and $\Phi_S$}
In this  section we discuss the finiteness question under the assumption
that we know the sign or the vanishing of $\Phi_S$; in particular we will
prove Lemma 2.5.2.C. On the other hand, in Sect.~4 the virial theorem will be
employed to obtain the required information on $\Phi_S$ from the EOS and the
LOM, independently of symmetry assumptions.

Finiteness in the radial direction, and the behaviour near the axis, 
 can be obtained rather easily from a qualitative discussion of the form of the
 potential. The key features of the potentials to be kept in mind here are:
\begin{enumerate}
\item $\phi_g \le 0$.
\item $\phi_c(r)$ behaves as in Definition 2.5.C and is monotonically decreasing
with $r$ from (\ref{phic}).
\item Inside the fluid region $\phi = \phi_g + \phi_c \le \Phi_S$,
and $\phi \nearrow \Phi_S$, as one approaches a locus of zero pressure 
 (irrespective of its location in space). 
\end{enumerate}

\begin{figure}[t!]
\begin{center}
\resizebox{!}{\textwidth}{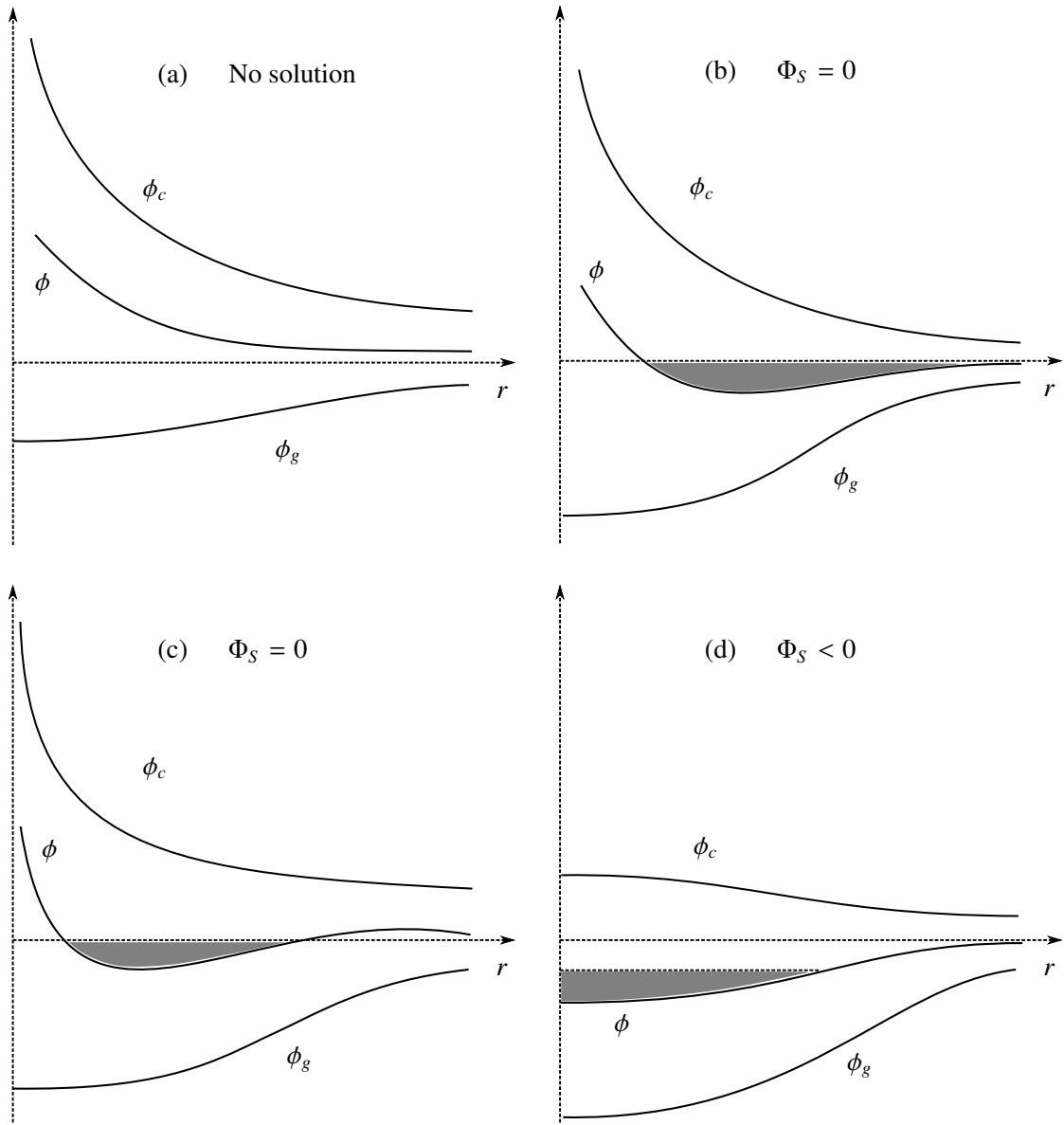}
\caption{Different a priori possible forms of the potentials $\phi_g$, $\phi_c$ and $\phi$.}
\end{center}
\end{figure}

Figures 1.(a--d) show qualitatively possible forms of the potentials together with the
respective fluid fillings (shaded), differing in their range with respect to infinity and the axis.
For positive potentials $\phi$ as in Fig.~1.(a) no solution exists, irrespective of the
value of $\Phi_S$, while the other figures show ranges for $\phi$ which are
a priori admissible but still without guarantee of existence of a solution. 
If the fluid is not axially symmetric, the curves are understood to be sections of
higher dimensional level surfaces of the potentials. 

While the behaviour of the fluid in the axis direction as included in Lemma
2.5.2.C is not seen from the above diagrams, it is a simple consequence of the 
Bernoulli equation. The Lemma is now proven as follows. 
 
\vspace{1ex} \noindent \textit{Proof of Lemma 2.5.2.C.}
By contradiction: Assume that the fluid is infinite in the $r$ direction. Then (\ref{ber}) and Definition 2.5.C yield
\begin{equation}
\label{limr}
\Phi_S = \lim_{r  \rightarrow \infty}\left( \phi + {\cal I} \right)  =  0,
\end{equation}
which contradicts the behaviour of $\Phi_S$, as claimed for finiteness in the $r$ direction. 
Here (\ref{limr}) in fact holds for the limit in any direction not parallel to the axis. 
The final assertion of the Lemma is obtained from (\ref{ber}), (\ref{phic}) and Definition 2.5.C.
as follows: 
Assuming that the fluid extends to infinity in the $z$ direction at $r$ we get
\begin{equation}
\label{limz}
\Phi_S = \lim_{z  \rightarrow \infty}\left( \phi + {\cal I} \right)  = \phi_c(r) =
\int_r^{\infty} \omega^2(r^\prime) r^\prime dr^\prime.
\end{equation} 
and hence $\Phi_S = 0$ implies $\omega = 0$.

\subsubsection*{Remark.}
As remarked after Definition 2.5, the condition in part C that 
$\lim_{r \rightarrow \infty} \phi_c(r) > -  \infty$ can be removed. 
In fact, if $\lim_{r \rightarrow \infty} \phi_c(r) =  - \infty$, the
statement of  Lemma 2.5.2.C remains correct and the proof is trivially adapted.

\section{General Results}

In Sect.~4.1 we give a result which, under rather restrictive conditions, forbids shells or ``hollow'' bodies. 
We continue in Sect.~4.2 with our main finiteness theorem, which rests on the suitably adapted virial
theorem, and we also obtain a more general bound on $\Phi_S$. 

\subsection{A ``no-shell'' result}
\textit{Proposition 4.1.} We consider a solution of (\ref{eos}--\ref{poi}) and assume part A of Definition 2.5, 
with $\rho_c \ge 0$. Then there cannot be any vacuum region not connected to infinity, i.e., the fluid ${\cal F}$
is not a ``shell'', and it is not ``hollow''.  
  
\vspace{1ex} \noindent \textit{Proof.} Assuming the contrary, ${\cal F}$ has an inner and an outer boundary, 
the latter either separating ${\cal F}$ from infinity or located at infinity, where $\phi$ takes on the 
value $\Phi_S$. We first apply the strong maximum principle, Thm.~9.6 of \cite{GT},
to  (\ref{poi}) on the region ${\cal R}$ consisting of ${\cal F}$ and the ``inner vacuum,'' but not the vacuum component 
connected to infinity. This implies that $\phi = \Phi_S$ is the maximum taken at the  boundary of ${\cal R}$
(the outer boundary of ${\cal F}$). But this maximum is also taken on at interior points of ${\cal R}$ , namely the inner boundary of 
${\cal F}$, and hence the maximum principle implies that  $\phi$ is constant on ${\cal R}$. The latter, however, is easily excluded 
from the assumptions.  

\vspace{1ex} \noindent \textit{Remark.} Clearly, this result does not exclude
toroidal rings of finite thickness, with or without central body.

\subsection{The finiteness theorem}

Our main result is now obtained by combining various Lemmas of the
preceding sections. The quantities $F$ and $D$ were defined in the
Introduction; the main conditions (\ref{ge}) and (\ref{le}) are discussed in
 remarks after the following Theorem and in Sect. 5.  \\

\textit{Theorem 4.2.} Assume we are given a solution of (\ref{eos}--\ref{ber}) such that
the EOS (\ref{eos}) satisfies the requirements of Lemma 2.3.1, and that $\rho_g \in W^{0,2}_{-3-\epsilon}$, $\epsilon > 0$ as in Lemma 2.4. 
Moreover, we require that $\phi_c$ or $\rho_c$ should satisfy one of A, B$_<$, B$_>$ or C in Definition 2.5 and that there holds in case  

\begin{description}
\item[A:]
one of 
\begin{eqnarray}
F(p) & \ge & 0 \quad \forall \, p \quad \mbox{and} \quad D(x) \ge 0 \quad \forall \, x,
\label{ge}\\
F(p) & \le & 0 \quad \forall \, p \quad \mbox{and} \quad D(x) \le 0 \quad \forall \, x,
\label{le}
\end{eqnarray}
\item[B$_<$:] 
$p \in W^{1,1}_{-4-\epsilon}$ and (\ref{ge}),
\item[B$_>$:] 
 $p \in W^{1,1}_{-4-\epsilon}$ and (\ref{le}),
\item[C:]  $p \in W^{1,1}_{-4-\epsilon}$ and (\ref{le}).  
\end{description}

\vspace{1ex} \noindent Then either $\Phi_S \ne 0$ and the fluid is finite, or $F \equiv 0$, $D \equiv 0$ and $\Phi_S = 0$.

\vspace{1ex} \noindent \textit{Corollary 4.2.1}. In case C above, assume that 
 (\ref{ge}) holds instead of (\ref{le}). Then the fluid is finite in all directions except possibly in the axial one.  

\vspace{1ex} \noindent \textit{Corollary 4.2.2}.
In case A above, assume that there hold (\ref{le}) and $\rho_c \ge 0$. Then either $\Phi_S < 0$ and the fluid is
finite, or $F \equiv 0$, $D \equiv 0$ and  $\Phi_S = 0$, and the fluid is infinite. 

\vspace{1ex} \noindent \textit{Proof.} We use the following modified version of the Pohozaev--Rellich identity
\cite{SP,FR}: Let $\vec{\xi} = x^i \frac{\partial}{\partial x^i}$ be a dilation, i.e., $\nabla_{(i} \xi_{j)} = 
\frac{1}{2}(\nabla_i \xi_j + \nabla_j \xi_i) = g_{ij}$, where $g_{ij}$ denotes the components of the 3-metric, 
and $\nabla_i$ is the covariant derivative with respect to $g_{ij}$. Then a simple calculation (compare~\cite{WS1,WS2}) shows that  
\begin{equation}
\fl \nabla_i\left[\left(\xi^j \nabla_j\phi_g + \frac{1}{2}\phi_g \right) \nabla^i \phi_g
- \frac{1}{2} \xi^i \nabla_j\phi_g \nabla^j \phi_g + 4\pi p \xi^i \right] =
2\pi \left[ \rho_g \left( \phi_g - 2 \xi^i \nabla_i \phi_c \right) + 6 p \right].  \label{poh}
\end{equation}
We first note that $p \in W^{1,q}_{-4-\epsilon}$, which in case A follows from Euler's equation (\ref{eul}). Next, by Gauss's law, 
the left hand side can be written as a surface term over a ball ${\cal S}$ of radius $R$. We now insert the forms of $\phi_g$, 
$\nabla \phi_g$, obtained in Lemma 2.4, and use the fact that $\psi_g$ and $\nabla_i \psi_g$
defined in this Lemma ``fall off faster'' than the leading 
terms, to show that the surface integral vanishes as $R \rightarrow \infty$. This follows
straightforwardly via Cauchy--Schwarz estimates and by applying 
the ``trace theorem'' (see e.g., Sect.~5.5 of~\cite{LE}): 
\begin{equation}
||  f |_{\partial S} ||_{L^2({\partial \mathcal{S}})} \le C ||f||_{ W^{1,q}(\mathcal{S})} 
\end{equation}   
(which holds for any function $f \in W^{1,q}(S)$, $q \ge 1$, and its extension
$f |_{\partial S}$ to $\partial S$), to the functions $\psi_g$, $\nabla_i \psi_g$ and
$p$. In the limit $R \rightarrow \infty$ we are left with
\begin{equation}    
 0  =  2\pi \int_{\mathbf R^3} \left[\rho_g \left(\phi_g   - 2 \xi^i \nabla_i \phi_c \right) + 6p \right] d\nu.
\end{equation}
Inserting (\ref{ber}) into the above expression gives 
\begin{equation}
\label{phis}
m \Phi_S = \int_{\mathbf R^3} \left[F(x) + 2 \rho_g D(x) \right] d\nu.
\end{equation}
The theorem is now obvious from the requirements and from Lemma 2.5.2.

\subsubsection*{Remarks.}
\begin{enumerate}
\item Note that the conclusion does not require falloff conditions for $F(x)$, and only the mild falloff conditions on 
$\phi_c(x^j)$ or $\rho_c$ from Definition 2.5. Even without such conditions, Eq. (\ref{phis}) and finiteness of 
$m\Phi_S$ imply that the integral on the right exists.
\item In the static case existence of finite, spherically 
symmetric solutions is known for polytropes with index $n \le 5$, which corresponds to $F \le 0$. On the other hand, if 
$F \ge 0$ solutions with finite mass do not exist except for the polytropes of index $n = 5$. This  suggests that the 
``realistic'' range for rotating fluids is given by conditions (\ref{le}). In fact the requirements of existence results 
for axisymmetric rotating fluids of which we are aware (\cite{LS} and the references therein) fall in this range, whereas
(\ref{ge})  might not allow for solutions at all.
\item If $\rho_c$ is at least weakly differentiable, the second term on the right in (\ref{phis}) can be rewritten as
\begin{equation}
 \int_{\mathbf R^3}  \rho_g D(x) d\nu =  
 \int_{\mathbf R^3}  \phi_g \left[x^i \frac{\partial}{\partial x^i} \rho_c + \frac{5}{2}\rho_c \right] d\nu + 
\mbox{surface terms},
\end{equation}
and, since $\phi_g \le 0$, the sign of the expression in the bracket determines (in)finite extent of the fluid 
region in the same manner as $D(x^i)$. The surface terms vanish in case A but
have to be handled with care in cases B$_<$, B$_>$ and C. 
\item To show the relation with the classical virial theorem (cf.~Sect.~1.4 of  \cite{GC}
or Sect.~2.8.1 of \cite{JT}) we use (\ref{poh}) and definition (\ref{centr_pot}) to obtain
(\ref{vir1}) below. To get (\ref{vir2})  which is the sum of the potential energy 
$E_{pot}= 1/2 \int_{\mathbf R^3}  \rho_g  \phi_g d\nu $, 
 the (bulk) kinetic energy 
$E_{kin} = 1/2 \int_{\mathbf R^3}  \rho_g  |U|^2 d\nu $,  
and the thermal energy (kinetic energy of the thermal motion)   
$E_{therm} =   3/2 \int_{\mathbf R^3} p d\nu $, 
one has to remove the second term in (\ref{vir1}) by partial integration using the continuity equation (\ref{con}), 
and assume that the velocity $\vec{U}$ falls off suitably in order for the surface
terms to vanish.
\begin{eqnarray}
\label{vir1}
0 &  = & 2\pi \int_{\mathbf R^3} \left\{ \rho_g \left[ \phi_g  - 2 U^j \nabla_j \left(\xi^i  U_i \right) +
2 |U|^2  \right] + 6p \right\} d\nu {} \\ 
\label{vir2}
& = & {} 4 \pi \left( E_{pot} + 2  E_{kin} + 2 E_{therm} \right).
\end{eqnarray}
\item In the static case $\Phi_S$ is the gravitational potential at the surface,
and it is related to the observed redshift. 
For a rotating, extended object, the redshift on the rim clearly depends on the velocity
of the rotation as well, but does not seem to bear any obvious relation to $\Phi_S$.   
\item Equation (\ref{phis}) is a formula for $\Phi_S$, but as such it requires a knowledge of
$F(x)$, as opposed to the mere sign condition on $F(p)$ used before. An estimate for $\Phi_S$
in terms of $F(p)$ and $D(x)$ is the following. 
\end{enumerate}

\vspace{1ex} \noindent \textit{Proposition 4.2.} Under the requirements of Lemma 2.3.2,
\begin{equation}
 \left| \Phi_S \right| \le  \sup_{\cal F} \left| \frac{F(p)}{\rho_g} \right| + 2 \sup_{\cal F} \left| D(x) \right|.
 \end{equation}

\vspace{1ex} \noindent \textit{Proof}. From (\ref{phis}),
\begin{eqnarray} 
m \left| \Phi_S \right| & \le & \int_{\mathbf R^3} \rho_g \sup \left|\frac{ F(x)}{\rho_g}  + 2 D(x)
\right| d\nu 
\le m \sup_{\cal F} \left|\frac{ F(\rho_g)}{\rho_g}  + 2 D(x) \right| \nonumber \\
& \le & m \sup_{\cal F} \left| \frac{F(p)}{\rho_g} \right| + 2 m \sup_{\cal F} \left| D(x)
\right|,
\end{eqnarray}
and the supremum of $ F(p)/\rho_g(p)$ exists, since from Lemma 2.3.2 $\lim_{p \rightarrow 0} F(p)/\rho_g(p) = \lim_{p \rightarrow 0} p/\rho_g(p)= 0$ .   

\section{Examples}
We examine here the limiting case of Theorem 4.2, and conclude with discussing ``power law rotations.'' 
Explicit examples are only available in the cylindrically symmetric case. 

\subsection{The limiting case}

The limiting case of Theorem 4.2 is $F(p) \equiv 0$, $D \equiv 0$ and $\Phi_S \equiv 0$.
The first condition yields the 1-parameter family of polytropic EOS (\ref{pol}) with index $n = 5$.
In the static case, the resulting PDE 
\begin{equation}
\Delta \phi_g  =  - 4\pi \rho_o \phi_g^5 
\end{equation}
has for each $\rho_o$ the well-known (cf.~e.g.~\cite{BT}) family of solutions
\begin{equation}
\phi_g  =  - \frac{m}{\sqrt{\frac{4\pi}{3} \rho_o m^4 + R^2}} 
\end{equation} 
parametrised by the mass $m$. All these solutions extend to infinity.

In the stationary case, if $\rho_c$ is differentiable, it follows from 
$0 = D = x^i \partial_i \phi_c + \frac{1}{2} \phi_c$ that $x^i \partial_i \rho_c + \frac{5}{2} \rho_c = 0$. 
This yields that $\phi_c$  and $\rho_c$  are homogeneous functions of degree $-1/2$ and $-5/2$, respectively.
That is to say, these functions have the form
\begin{equation}
\phi_c = z^{-\frac{1}{2}}\sigma \left(\frac{x}{z},\frac{y}{z} \right),  \qquad
\rho_c = z^{-\frac{5}{2}}\tau \left(\frac{x}{z},\frac{y}{z} \right), \qquad z \neq 0 
\label{sip}
\end{equation}
for some arbitrary (but mutually related) functions $\sigma(x,y)$ and  $\tau(x,y)$.
To determine the solution, we have to solve
\begin{equation}
\label{delphi}
\Delta \phi = \Delta \left(\phi_g + \phi_c \right) = - 4\pi \rho_o \phi^5 +
4 \pi z^{-\frac{5}{2}}\tau \left( \frac{x}{z},\frac{y}{z} \right)    
\end{equation}
in ${\cal F}$, and $\Delta \phi = 0$ in ${\cal V}$ (if present).
We remark that (\ref{delphi}) is scale invariant under
\begin{equation} 
\phi(x^i) \longrightarrow \sqrt{\frac{1}{\lambda}} \phi \left( \frac{x^i}{\lambda} \right), \quad \forall
~ \lambda = \mathrm{const.} > 0.
\label{sc}
\end{equation} 

Thus the motion determined by $D \equiv 0$ could be called ``scale invariantly rotating polytrope of index 5.'' 

Scale invariance is sometimes useful for getting information about (non-)existence of solutions,
in particular in combination with the scaling behaviour of the energy functional. We do not go into details here. 

Under restriction to axially symmetric rotation laws 
(about the the $z$ axis)  (\ref{sip}) becomes
\begin{equation}
\label{axs}
\phi_c = z^{-\frac{1}{2}}\alpha \left(\frac{r}{z} \right),  \qquad
\rho_c = z^{-\frac{5}{2}}\beta \left(\frac{r}{z} \right), \qquad z \neq 0 
\end{equation}
for some arbitrary (again related) functions $\alpha(r)$ and  $\beta(r)$. 
However, by Lemma 3.1, the rotation law must be cylindrically symmetric. 
Choosing $\alpha$ and $\beta$ in (\ref{axs}) appropriately, we have
\begin{equation}
\label{cyls}
\phi_c(r) = 2 \frac{C^2}{\sqrt{r}}, \qquad \rho_c(r) = \frac{C^2}{8\pi r^{\frac{5}{2}}},   \qquad 
C = \mathrm{const.} \quad \mbox{in}~ {\cal F}.
\end{equation}
 We obtain the following behaviour for the ``scale invariant rotation'' $\omega_{si}(r)$,
 whose falloff interestingly lies between the Kepler angular velocity $\omega_{\odot}$ 
(actually Copernicus \cite{CO}, as this is for circular orbits) and the observed galaxy rotation curves $\omega_{@}$: 
  \begin{equation}
\omega_{\odot} =  \frac{C}{r^{\frac{3}{2}}}, \qquad
 \omega_{si} =  \frac{C}{r^{\frac{5}{4}}}, \qquad 
  \omega_{@} \sim  \frac{C}{r}.
 \end{equation}  

Regarding finiteness, we note that $\omega$ and $\phi_c$ are singular on the axis,
whence Theorem 4.2.A is not applicable; in particular the maximum of $\phi$
is obviously taken at the axis. However, (\ref{cyls}) implies that 
$\phi =  \phi_c + \phi_g =  2 C^2/\sqrt{r}  - m/R + o(1/R^{1 + \epsilon}) \ge 0$
for sufficiently large $r$. Hence
if a solution exists, the form of $\phi$ must be as in Fig.~1.(c), 
while the form of Fig.~1.(b) is excluded at least in the cylindrically symmetric case
considered here.  Thus the fluid is either finite (with a hole near the
axis) or it is infinite in the z-direction (respecting Proposition 3.1).
The former case would probably give a toroidal configuration.

  We can now discuss the consequences of the above arguments for 
``power law rotations.''

\subsection{Power law rotation}  
\vspace{1ex} \noindent \textit{Proposition 5.2.}
 Assume (\ref{eos}--\ref{ber}) have a solution which satisfies the conditions of
 Theorem 4.2.C with the rotation law 
\begin{equation}
\label{pow}
\phi_c(r) = \frac{C^2}{2(k-1) r^{2 (k-1)}} ~~\Longleftrightarrow ~~
\omega(r) = \frac{C}{r^k}, \qquad k,C \in \mathbf{R}.
\end{equation}
Then depending on the value of $k$ the fluid has the following properties regarding infinity and the axis:
\begin{enumerate}
\item $k < \frac{3}{2}$: It is finite in the radial direction.
\item $k > 1$: It has a ``hole'' near the axis. 
\item $k \ge \frac{5}{4}$, $F \le 0$ and  $p \in W^{1,1}_{-4-\epsilon}$:
It is either finite (with a hole near the axis), or infinite in the axis direction (respecting Proposition 3.1).
\end{enumerate}
Moreover, if the solution is finite in the axis direction in case (iii), the
conclusion (finiteness) holds without the falloff condition on the pressure.
     
\vspace{1ex} \noindent \textit{Proof.}
For $k \le 1$ statement (i) follows form the fact that $\phi_c$ diverges at infinity. 
For $1 < k < 3/2$ the same conclusion can be inferred from the argument used in Sec.~5.1 
to show that for the ``scale invariantly'' rotating polytrope $\phi = \phi_c + \phi_g \geq 0$ for sufficiently large $r$. 
Conclusion (ii) is obvious from the fact that $\phi_c$ diverges at the axis.
To see (iii) we calculate from Definition (\ref{D})  
\begin{equation}
 \quad D(r) = \frac{C^2 (5 - 4k)}{4(k-1)r^{2(k-1)}},
\end{equation}
and use Theorem 4.2.C for the generic case, and the discussion of Sect. 5.1 for the limiting case 
with $\Phi_S = 0$. For the final statement, we note that (\ref{eul}) with Lemma
2.4 and $k \ge 5/4$ imply that $p = O(r^{-7/2})$. This is sufficient in
order for the surface term in the integral of (\ref{poh}) to  vanish as $R \rightarrow \infty$, provided the fluid is 
finite in the axis direction.

\ack
We are grateful to Piotr Bizo\'{n} for pointing out the scaling behaviour
(\ref{sc}) of the liming case, and
we also acknowledge helpful discussions with Lars Andersson, Robert Beig,
Piotr T.~Chru\'{s}ciel, J.~Mark Heinzle, Edward Malec, Reinhard Meinel, Herbert Pfister, Roland Steinbauer and Claes Uggla. 
Moreover, we thank the referees for useful comments.

W.S.~acknowledges support from \"Osterreichische Forschungsgemeinschaft (project MOEL 443)
and from The Spanish Ministerio de Ciencia e Innovacion (project
FIS2009-07238).


\section*{References}
\begin{thebibliography}{10}
\bibitem{AK} Arnold V I and Khesin B A 1998 {\it Topological Methods in
Hydrodynamics} (New York, Berlin, Heidelberg: Springer) 
\bibitem{SC} Chandrasekhar S 1958 {\it An introduction to the study of
Stellar Structure} (New York: Dover)
\bibitem{JT} Tassoul J 2000 {\it Stellar rotation} (Cambridge: Cambridge University Press)
\bibitem{BT} Binney J and Tremaine S 1987 {\it Galactic Dynamics} (Princeton: Princeton University Press) Sect. 2.2. 
\bibitem{WS1} Simon W 1993 {\it Class Quantum Grav.} {\bf 10} 177
\bibitem{LM} Lindblom L and Masood-ul-alam A K M in: 1993 {\it Directions in General
Relativity. Proceedings of the 1993 International Symposium, Maryland, Vol.2}
eds. Hu B L and Jacobson T A (Cambridge: Cambridge University Press) 
\bibitem{WS2} Simon W 2002 in: {\it The Conformal Structure of Space-Time;
Lecture Notes in Physics 604}, Frauendiener J and Friedrich H {\it eds} (Berlin: Springer) Sect. 11
\bibitem{MH} Heinzle J M 2002 {\it Class. Quantum Grav.} {\bf 19} 2835  
\bibitem{HU} Heinzle J M and Uggla C 2003 {\it Ann. Phys.} {\bf 308}, 18
\bibitem{AB} Auchmuty J F G and Beals R 1971  {\it Arch. Rat. Mech. Anal.} {\bf 43} 255
\bibitem{LS} Luo T and Smoller J 2009 {\it Arch. Rat. Mech. Anal.} {\bf 191} 447
\bibitem{JBK} Jalocha J, Bratek L and Kutschera M 2008 {\it Astrophys. J}
{\bf 679} 373
\bibitem{DS}  Saari D 2011 {\it Dark matter: is it really a problem?} talk,
University of Vienna
\bibitem{RR} Rein G and Rendall A 2000 {\it Math. Proc. Camb. Phil. Soc.} {\bf
128} 363 
\bibitem{GC} Collins II G W 1978 {\it The virial theorem in stellar astrophysics}
(Tucson: Pachart Publishing House).
\bibitem{SBEG} Bonazzola S, Gourgoulhon E 1994 {\it Class. Quantum Grav.} {\bf 11}
1775  
\bibitem{BS} Beig R and Simon W 1992 {\it Commun. Math. Phys.} {\bf 144} 373
\bibitem{MO} McOwen R 1979 {\it Comm. Pure Appl. Math.} {\bf 32} 783
\bibitem{RB} Bartnik R 1986 {\it Comm. Pure Appl. Math.} {\bf 39} 661
\bibitem{GT} Gilbarg D and Trudinger N S 1998 {\it Elliptic Partial Differential Equations of Second Order}
(Berlin: Springer) 
\bibitem{LL} Lindblom L 1992 {\it Phil. Trans. Roy. Soc. London A} {\bf 340}, 353 
\bibitem{BG} Guilfoyle B 2005 {\it Class. Quantum Grav.} {\bf 22}, 1599
\bibitem{LE} Evans L 1998 {\it Partial Differential Equations} (Providence: American Mathematical Society) 
\bibitem{SP} Pohozaev S 1965 {\it Sov. Math. Doklady} {\bf 6} 1408
\bibitem{FR} Rellich F 1940 {\it Math. Zeitschrift} {\bf 46} 635
\bibitem{CO} Copernicus N 1543 \textit{De revolutionibus orbium coelestium} (Norimberga: Iohannes Petreius)

\end{thebibliography}
\end{document}